\documentclass[runningheads]{llncs}

\usepackage[T1]{fontenc}
\usepackage{amsmath,amssymb}
\usepackage{graphicx}
\usepackage{booktabs}
\usepackage{xcolor}
\usepackage{mathtools}
\usepackage{enumitem}
\usepackage{hyperref}
\hypersetup{colorlinks=true,linkcolor=blue,citecolor=blue,urlcolor=blue}

\newcommand{\VECK}{\mathsf{VECK_{EL}}}
\newcommand{\VECKplus}{\mathsf{VECK^+_{EL}}}
\newcommand{\VECKstar}{\mathsf{VECK^{\star}_{EL}}}
\newcommand{\F}{\mathbb{F}}

\newcommand{\textbfmath}[1]{{\bfseries\boldmath #1}}

\title{Plaintext-Scale Fair Data Exchange}

\author{Majid Khabbazian}
\authorrunning{}
\institute{University of Alberta, Canada}

\begin{document}
\maketitle

\begin{abstract}
The Fair Data Exchange (FDE) protocol (CCS’24) achieves atomic, pay‑per‑file exchange with a constant on‑chain footprint, but existing implementations do not scale: proof verification, for instance, can take hours even for files of only tens of megabytes.

In this work, we present two FDE implementations: \(\VECKplus\) and \(\VECKstar\). \(\VECKplus\) reduces client‑side verification to \(O(\lambda)\)---independent of file size---where \(\lambda\) is the security parameter.
Concretely, \(\VECKplus\) brings verification time to \(\approx 1\,\mathrm{s}\) on a commodity desktop for any file size.

\(\VECKplus\) also significantly reduces proof generation time by limiting expensive range proofs to a \(\Theta(\lambda)\)-sized subset of the file. This improvement is especially beneficial for large files, even though proof generation and encryption are already precomputable and highly parallelizable on the server: for a \(32\,\mathrm{MiB}\) file, for instance, proof generation time drops from \(\approx 6{,}295\,\mathrm{s}\) to \(\approx 4.8\,\mathrm{s}\) (\(\approx 1{,}300\times\) speed-up).

As in the existing ElGamal implementation, however, \(\VECKplus\) retains exponential ElGamal over the full file. Consequently, the client must perform ElGamal decryption and download ciphertexts that are \(\gtrsim 10\times\) the plaintext size. We address both drawbacks in the second implementation, \(\VECKstar\): we replace bulk ElGamal encryption with a fast, hash-derived mask and confine public-key work to a \(\Theta(\lambda)\) sample tied together with a file-size-independent zk-SNARK, adding \(<0.1\,\mathrm{s}\) to verification in our prototype. Importantly, this also reduces the communication overhead from \(\gtrsim 10\times\) to \(<50\%\). Together, these changes yield \emph{plaintext-scale} performance.

Finally, we bridge Bitcoin’s secp256k1 and BLS12‑381 with a file-size-independent zk‑SNARK to run FDE fully off‑chain over the Lightning Network, reducing fees from $\approx \$10$ to \(<\!\$0.01\) and payment latency to a few seconds.
\end{abstract}

\section{Introduction}

    Digital commerce increasingly revolves around the one-shot sale of large digital assets---scientific data sets, proprietary machine-learning weights, high-resolution media, even genomic archives---between parties with no prior trust relationship. Ensuring that the buyer actually receives the promised file while the seller simultaneously receives payment is therefore a foundation stone of the data-driven economy. Traditional escrow or licensing services resolve this tension by inserting costly legal or institutional intermediaries; blockchain-based atomic fair-exchange protocols aim to provide the same fairness guarantee with only a smart contract as referee. Achieving that goal without inflating on-chain fees or forcing the parties through many interactive rounds, however, remains challenging once file sizes grow into the multi-megabyte range.

Prominent blockchain-based fair exchange protocols---FairSwap~\cite{DziembowskiEF18}, \linebreak FileBounty~\cite{JaninQMG20}, and FairDownload~\cite{He00WW21}---follow a dispute-driven model in which misbehavior is resolved through explicit on-chain arbitration. FairSwap requires both parties to remain online during a complaint window, with buyers providing Merkle proofs of misbehavior. FileBounty supports dispute resolution at any point using zk-SNARKs, but assumes partial downloads have proportional utility. FairDownload uses off-chain exchange of signed chunks and resolves disputes via $O(\log k)$ Merkle proofs.

The FDE protocol~\cite{TasSZMKBN24} eliminates dispute timers and
achieves constant-size on-chain communication. The seller posts a KZG
polynomial commitment~\cite{KateZG10} to the data together with a
constant-size \emph{Verifiable Encryption of Committed Knowledge}
(VECK) proof showing that the ciphertexts sent to the buyer correctly
encrypt the committed evaluations. The buyer verifies this proof,
escrows the payment on-chain, and later decrypts using a key posted by the seller—without any dispute phase.

\color{black}
A main target application is Ethereum’s ProtoDanksharding
data-availability layer. After blob data expires from the L1 chain,
archival nodes may continue storing it while the chain retains only a short KZG commitment~\cite{TasSZMKBN24}. A user who later wishes
to recover the blob runs an FDE instance with such a node: the node
uses VECK to prove that its encrypted file is consistent with the commitment, the
user escrows payment on-chain, and the node is paid only upon
revealing the decryption key, yielding a trust-minimized market for
historical blob data.
\color{black}

A primary obstacle to practical FDE deployment is computational overhead. 
In the implementation of~\cite{TasSZMKBN24}, verifying the 
ElGamal-based VECK proof for a \(128\,\mathrm{KiB}\) file (i.e., \(4096\) BLS12-381 field 
elements) takes about \(34.15\,\mathrm{s}\). On our hardware, the same verification 
completes in roughly \(8.84\,\mathrm{s}\). Over the range of file sizes we tested, 
verification time grows essentially linearly in the number of elements; under 
this scaling, verification would take on the order of \(2.5\) hours for a 
\(128\,\mathrm{MiB}\) file and about \(20\) hours for a \(1\,\mathrm{GiB}\) file.

Communication overhead further compounds the challenge: the ElGamal-based FDE implementation expands transmitted data by about \(10\times\) relative to plaintext, which is prohibitive for large-file transfers~\cite{TasSZMKBN24}.

\textbf{Our contributions.}
We make FDE practical for large files by proposing two implementations:

\emph{(1) Verification in \(O(\lambda)\).}
Our first implementation reduces client-side verification to \(O(\lambda)\), independent of file size, bringing verification down to about one second on commodity hardware. It also substantially lowers prover work, but it \emph{retains} exponential ElGamal encryption: encryption time is unchanged and bandwidth remains about \(10\times\). The upside is that proof generation is server-side, precomputable, and parallelizable across cores.

\emph{(2) Near-plaintext exchange.}
Our second implementation replaces expensive ElGamal encryption/decryption with a lightweight hash-derived mask, using public-key operations only on a \(\Theta(\lambda)\) sample tied together with a file-size-independent zk-SNARK. This not only makes encryption/decryption extremely fast, but also eliminates the bandwidth bloat from ElGamal. The cost is a fixed, file-size-independent overhead of about \(0.1\,\mathrm{s}\) in verifier time and about \(10\,\mathrm{s}\) in prover time (on 8 cores) due to the zk-SNARK.

Beyond the above, we resolve the secp256k1\(\leftrightarrow\)BLS12-381 mismatch with a compact, file-size-independent zk-SNARK that bridges Bitcoin adaptor signatures to our proofs. This enables fully off-chain execution over the Lightning Network, cutting fees from \(\sim\$10\) to \(<\$0.01\) and latency from tens of minutes to \(\approx 1\,\mathrm{s}\).


\section{Background}
\label{sec:background}

\subsection{Notation and Definitions}

    Let $\lambda \in \mathbb{N}$ denote the security parameter. A non-negative function $\sigma(\lambda)$ is said to be \emph{negligible} if, for every polynomial $p(\lambda)$, there exists a sufficiently large $\lambda_0$ such that for all $\lambda \geq \lambda_0$:
    $\sigma(\lambda) \leq \frac{1}{p(\lambda)}$.

    For a random variable $x$, we write $x \gets_{R} X$ to indicate that $x$ is drawn uniformly at random from the set $X$. Throughout this paper, we denote by $\mathbb{F}_p$ the finite field of prime order $p$. We let $\mathbb{F}_p[X]$ represent the set of all univariate polynomials with coefficients in $\mathbb{F}_p$.

    We work with elliptic-curve groups \(G_1\), \(G_2\), and a target group \(G_T\), each of prime order \(p\), and a bilinear pairing \(e: G_1 \times G_2 \to G_T\) with the usual properties. We instantiate these groups on the pairing-friendly curves BLS12-381 and BLS12-377; the corresponding subgroup orders are \(\approx 255\)-bit and 253-bit primes, respectively. Both target the \(\approx 128\)-bit security level.


    For a set $S$ and a function $\phi(X)\in \mathbb{F}_p[X]$, $\phi_S(X)$ denotes the minimal-degree polynomial in $\mathbb{F}_p[X]$ satisfying $\phi_S(i) = \phi(i)$ for all $i \in S$. Additionally, define
    \[
        V_S(X) \coloneqq \prod_{i \in S}(X - i).
    \]

\subsection{KZG Polynomial Commitments}

The KZG polynomial commitment scheme~\cite{KateZG10} provides a cryptographic method for succinctly committing to polynomials and efficiently generating evaluation proofs. It leverages pairing-based cryptography to achieve constant-size commitments and evaluation proofs, independently of the polynomial degree.
Formally, KZG consists of the following algorithms:

\begin{itemize}

    \item $\mathsf{Setup}(1^\lambda, n)\rightarrow \mathsf{crs}$:
    generates $\mathcal{G}$: elliptic-curve groups $G_1, G_2, G_T$ of prime order $p \geq 2^{2\lambda}$, with generators $g_1 \in G_1$, $g_2 \in G_2$, $g_T \in G_T$, and a bilinear pairing map $e : G_1 \times G_2 \rightarrow G_T$. 
        Samples a uniformly random secret $\tau\gets_R\mathbb{F}_p$ and publish
        \[
          \mathsf{crs}=\bigl(\mathcal{G},\{g_1^{\tau^i}\}_{i=1}^{n},\{g_2^{\tau^i}\}_{i=1}^{n}\bigr).
        \]

    \item $\mathsf{Commit(\mathsf{crs},\phi)}\rightarrow C$:  
        Given the $\mathsf{crs}$ and the coefficients of $\phi(X)\in \mathbb{F}_p[X]$, output the commitment
        $C = g_{1}^{\phi(\tau)}$.

    \item $\mathsf{Open}(\mathsf{crs}, \phi, i)\rightarrow \pi $: Given public parameters $\mathsf{crs}$, polynomial $\phi(X)\in \mathbb{F}_p[X]$, and an evaluation point $i$, the prover outputs the opening proof:
    \[
    \pi = g_1^{\frac{\phi(\tau)-\phi(i)}{\tau - i}} \in G_1.
    \]

    \item $\mathsf{Verify}(\mathsf{crs}, C_\phi, i, \phi(i), \pi)\rightarrow 0/1$: To verify an opening, the verifier checks the pairing equation:
    \[
    e(C_\phi/g_1^{\phi(i)}, g_2) \stackrel{?}{=} e(\pi_i, g_2^{\tau - i}),
    \]
    and outputs 1 if it holds, otherwise 0.

    \item $\mathsf{batchOpen}(\mathsf{crs}, \phi, S)\rightarrow \pi$: Given public parameters $\mathsf{crs}$, polynomial $\phi(X)$, and multiple distinct evaluation points $S$, the prover computes the evaluations $\{\phi(i)\}_{i\in S}$ and produces a single aggregated proof:
    $\pi = g_1^{q(\tau)}$,
    where
    \[
    q(X)=\frac{\phi(X)-\phi_S(X)}{V_S(X)}.
    \]

    \item $\mathsf{batchVerify}(\mathsf{crs}, C_\phi, S, \{\phi(i)\}_{i\in S}, \pi)\rightarrow 0/1$: The verifier efficiently checks the aggregated proof $\pi$ by verifying the following pairing equation:
    \[
    e\left(C_\phi / g_1^{\phi_S(\tau)},\, g_2\right) \stackrel{?}{=} e\left(\pi,\, g_2^{V_S(\tau)}\right).
    \]
    The verifier outputs 1 if this equation holds and 0 otherwise.

\end{itemize}

\subsection{FDE Protocol}
    FDE is a blockchain-based protocol which enables a storage server and a client to atomically exchange data for payment. Atomicity guarantees that the server receives payment if and only if the client obtains the promised data. The FDE protocol leverages the KZG polynomial commitment scheme, chosen due to its constant-size commitments and efficient batchable opening proofs, making it particularly suitable for scenarios where clients may retrieve subsets of data. Additionally, KZG commitments are already widely adopted in blockchain ecosystems, notably in Ethereum's Danksharding for data availability, making FDE naturally compatible with existing infrastructure.

    In the FDE protocol, data is represented as evaluations of a polynomial $\phi(\cdot)$ of degree $\ell\leq n$, that is as $\{\phi(i)\}_{i=0}^{\ell}$. The server first publishes a public verification key~$\mathsf{vk}$ to a blockchain smart contract, alongside specific transaction details such as the agreed price and the client’s blockchain address (step 1). Off-chain, the server then sends the encrypted evaluations $\{\mathsf{ct}_i\}_{i=0}^{\ell}$ of each data point $\{\phi(i)\}_{i=0}^{\ell}$ to the client. These ciphertexts are accompanied by a cryptographic proof showing that each ciphertext $\mathsf{ct}_i$ correctly encrypts the corresponding polynomial evaluation $\phi(i)$ committed to by a KZG polynomial commitment $C_{\phi}$, under a secret decryption key $\mathsf{sk}$ that matches the previously submitted verification key~$\mathsf{vk}$ (step~2).
    
    Upon receiving and verifying these ciphertexts and associated proofs, the client locks the agreed-upon funds in the blockchain smart contract (step 3). The server subsequently can claim these funds only by revealing the correct decryption key~$\mathsf{sk}$ that corresponds to the public verification key~$\mathsf{vk}$ (step 4). After the server publishes the decryption key, the client retrieves it from the blockchain (step 5), allowing immediate decryption of the ciphertexts and recovery of the original committed data (step 6). If the server fails to reveal the secret key within a specified timeout, the client recovers the funds locked in the contract, thus preserving atomicity.
    
    FDE satisfies three critical properties: correctness (honest parties always succeed), client-fairness (the server obtains payment only if the client receives the data), and server-fairness (the client learns nothing about the data unless the server is paid).

\subsection{Verifiable Encryption under Committed Key (VECK)}

At the heart of the FDE protocol lies a novel cryptographic primitive called VECK. At a high level, VECK enables a prover (in our context, the server) to demonstrate that a set of ciphertexts indeed encrypts evaluations of a polynomial at specific points, using an encryption key consistent with a publicly committed verification key. Concretely, VECK allows efficient verification of ciphertext correctness against a polynomial commitment without revealing the underlying plaintext evaluations or the encryption key itself. This construction ensures that the verifier (client) learns no additional information beyond the correctness of ciphertexts until the prover reveals the corresponding decryption key.

\textbf{Formal description.} Let $(\mathsf{Setup}, \mathsf{Commit})$ be a non-interactive binding commitment scheme, where $\mathsf{Setup}(1^\lambda, n)\rightarrow\mathsf{crs}$ generates a public common reference string, and $\mathsf{Commit}(\mathsf{crs}, w\in W)\rightarrow C_w$ generates a commitment to $w$. A non-interactive VECK scheme for a class of functions $\mathcal{F} = \{F: W \rightarrow V\}$ is defined as follows~\cite{TasSZMKBN24}:

\begin{itemize}
    \item $\mathsf{VECK.Gen(\mathsf{crs})} \rightarrow \mathsf{pp}$: A PPT algorithm that, given the $\mathsf{crs}$, outputs parameters $\mathsf{pp}$ and defines relevant spaces. The parameters $\mathsf{pp}$ are implicitly provided to subsequent algorithms.

    \item $\mathsf{VECK.Enc(F, C_w, w)} \rightarrow (\mathsf{vk}, \mathsf{sk}, \mathsf{ct}, \pi)$: A PPT algorithm run by the server that takes $\mathsf{(F, C_w, w)}$ and outputs a verification key $\mathsf{vk}$, a decryption key $\mathsf{sk}$ and the encryptions $\mathsf{ct}$ of $F(w)$, as well as a proof $\pi$.

    \item $\mathsf{VECK.Ver_{ct}(F, C_w, vk, ct, \pi)} \rightarrow \{0,1\}$: A deterministic polynomial-time algorithm executed by the client that returns either \emph{accept} or \emph{reject}.

    \item $\mathsf{VECK.Ver_{key}(vk, sk)} \rightarrow \{0,1\}$: A deterministic polynomial-time algorithm executed by the blockchain or a trusted third party to verify the validity of the secret key.

    \item $\mathsf{VECK.Dec(\mathsf{sk}, \mathsf{ct})} \rightarrow v/\bot$: A deterministic polynomial-time algorithm executed by the client that outputs a value in $V$ or a failure symbol $\bot$.
\end{itemize}

\color{black}
\textbf{Security notions.}
Following Tas et al.~\cite{TasSZMKBN24}, we say that a VECK scheme is secure if it
satisfies \emph{correctness}, \emph{soundness}, and \emph{(computational) zero-knowledge}
as defined there (Definition~3.1).
Informally, correctness requires that honestly generated ciphertexts and
proofs always verify and decrypt to $F(w)$; soundness rules out any PPT
adversary that can produce accepting ciphertexts whose decryption is
inconsistent with the committed data; and computational zero-knowledge
guarantees that the ciphertext, verification key, and proof reveal no
additional information about $w$ beyond $F(w)$.
In Theorems~\ref{thm:veckplus-security} and~\ref{thm:veckstar-security} we
show that our concrete constructions $\VECKplus$ and $\VECKstar$ satisfy these
notions, and hence are secure VECK instantiations in the sense of
Tas et al.~\cite{TasSZMKBN24}.
\color{black}


\color{black}
\subsection{Reed--Solomon Codes: High-Level Intuition}\label{subsec:rs}

We briefly recall the coding-theoretic intuition underlying our constructions and
refer to Appendix~\ref{app:rs-background} for a more formal treatment.

Let $\phi(X) \in \F_p[X]$ be a polynomial of degree at most $\ell$ encoding the
file. A basic fact is that $\phi$ is uniquely determined by its evaluations
at any $\ell+1$ distinct points: given $\{\phi(i)\}_{i \in T}$ for a set
$T$ of size $|T| = \ell+1$, one can efficiently interpolate $\phi$.

Reed--Solomon (RS) codes formalize the case where we evaluate $\phi$ on a larger
domain. Fix $m \ge \ell$ and consider the evaluation points $[m] = \{0,1,\dots,m\}$.
The associated RS codeword is the vector
\[
  \bigl(\phi(0),\phi(1),\dots,\phi(m)\bigr) \in \F_p^{\,m+1}.
\]
Standard RS decoding algorithms can still efficiently recover $\phi$ from these
$m+1$ symbols even if up to
\[
  t \;\le\; \bigl\lfloor (m - \ell)/2 \bigr\rfloor
\]
entries are arbitrarily corrupted (and we do not know which ones).

In practice, it is cheaper to first \emph{detect} whether any corruption occurred
and to invoke full RS decoding only when necessary. This detect-then-correct
pattern is exactly what we exploit later in our constructions.

For fixed code parameters $(\ell,m,t)$ as above, we model RS
decoding via two black-box algorithms. The decoder
$
  \mathsf{RS.Dec} : \F_p^{\,m+1} \to \F_p[X]_{\le \ell} \cup \{\bot\}
$
is any deterministic RS decoder that, on input a word $y$, returns the
(unique) degree-$\le\ell$ polynomial within Hamming distance at most $t$
of $y$, if it exists, and $\bot$ otherwise. The detector
$\mathsf{RS.Det} : \F_p^{\,m+1} \to \{0,1\}$ is defined by $\mathsf{RS.Det}(y) := \mathbf{1}[\mathsf{RS.Dec}(y)\neq\bot]$.
In the VECK constructions we simply treat $\mathsf{RS.Det}$ and $\mathsf{RS.Dec}$ as such
black-box subroutines.

\color{black}

\color{black}
\section{High-Level Overview of Our VECK Constructions}

Recall from Tas et al.~\cite{TasSZMKBN24} that VECK provides a way for a
server to prove that a ciphertext vector encrypts evaluations of a committed
polynomial, under a decryption key whose verification key has been posted to
the blockchain. Their ElGamal-based VECK instantiation already gives a
constant-size proof, but both the public-key work and the ciphertext size scale
linearly with the file length: every polynomial evaluation is encrypted under
exponential ElGamal and participates in the proof. As a result, client-side
verification is linear in the number of data points and the transmitted data is
roughly $10\times$ the plaintext size.

Our first construction, $\VECKplus$, keeps the same ElGamal-based VECK
building block, but uses Reed–Solomon (RS) coding to cap the verifier’s work at
$O(\lambda)$, independent of file size. Intuitively, instead of treating the
$\ell+1$ committed evaluations as a flat vector, we first view them as part of a
longer RS codeword of length $m \approx \beta(\ell+1)$. The server produces an
ElGamal-based VECK ciphertext for every position of this codeword, and the
verifier checks correctness only on a small, Fiat–Shamir-derived random sample
$S_R \subseteq [m]$ of size $R = \Theta(\lambda)$. If all sampled positions are
correct, the RS code guarantees (except with negligible probability in
$\lambda$) that the entire codeword is consistent with a low-degree polynomial,
so the client can decode and safely recover the desired evaluations. Thus
$\VECKplus$ trades a modest RS redundancy for verification time that is
essentially flat in the file size, while inheriting correctness, soundness, and
zero-knowledge from the underlying ElGamal-based VECK.

Our second construction, $\VECKstar$, aims to remove the remaining
bottlenecks of ElGamal over the full file: slow public-key decryption and
$\approx 10\times$ bandwidth blow-up. Here the core idea is to encrypt the file
\emph{symmetrically} under a one-time, hash-derived mask, and use public-key
crypto only on a $\Theta(\lambda)$-sized sample. Concretely, the server samples a
secret key $\mathsf{sk}$, derives a pseudorandom mask $\{H(\mathsf{sk},i)\}_{i\in[m]}$,
and sends the masked plaintext $\{ct_i := \phi(i) + H(\mathsf{sk},i)\}$ as the
bulk ciphertext. To convince the client that this mask is well formed and tied
to the posted verification key $\mathsf{vk} = h^{\mathsf{sk}}$, the server (i) runs the
ElGamal-based VECK of~\cite{TasSZMKBN24} on a random sample $S_R$ of
positions, and (ii) produces a short zk-SNARK showing that, on those sampled
indices, the masked values and the ElGamal ciphertexts encrypt the same
underlying evaluations under the same key. The client then only needs to
verify the ElGamal-based VECK on $|S_R| = \Theta(\lambda)$ positions plus a
single SNARK proof, both of which are independent of the file length. At the
same time, encryption and decryption on the full file are symmetric-key fast,
and the overall ciphertext size is driven by the masked plaintext itself, leading
to near-plaintext communication overhead.
\color{black}

\section{Proposed FDE Implementations}\label{sec:proposed-fde}

Our primary goal is to make FDE efficient and practical for large-scale files. To achieve this, we propose two alternative implementations of the FDE protocol: the first leverages RS codes to substantially reduce proof complexity and runtime, while the second further enhances computational efficiency and minimizes communication overhead through optimized masking techniques combined with file-size-independent zk-SNARK proofs.

These implementations rely on two novel ElGamal-based VECK protocols, denoted as $\VECKplus$ and $\VECKstar$. Before detailing these implementations, we briefly introduce notation and definitions common to both implementations.

\subsection{Common Notation and Definitions}

As in~\cite{TasSZMKBN24}, we use the KZG polynomial commitment scheme, and define the function $F$ in VECK as the polynomial evaluation at a given set of points.

Our first proposed construction, $\VECKplus$, uses ElGamal-based encryption. We choose ElGamal primarily because it achieves significantly lower bandwidth compared to the Paillier-based VECK from~\cite{TasSZMKBN24}, despite its relatively higher proof-time complexity. This trade-off is justified in $\VECKplus$, as our design already reduces proof time, making the bandwidth savings more impactful.

Our second construction, $\VECKstar$, also employs ElGamal-based encryption, primarily to maintain consistency between the two implementations. However, unlike in $\VECKplus$, the bandwidth advantage of ElGamal over Paillier is less critical, as $\VECKstar$ already drastically reduces communication overhead through optimized masking and zk-SNARK integration. Thus, in the context of $\VECKstar$, one might prefer Paillier encryption to leverage its slightly better computational efficiency, as reported in~\cite{TasSZMKBN24}.

Let $\mathsf{VECK_{El}}$ denote the Elgamal-based VECK protocol proposed in~\cite{TasSZMKBN24}.
Since our VECK implementations have similarities with $\mathsf{VECK_{El}}$, instead of constructing them from scratch, we strive to call $\mathsf{VECK_{El}}$ operation whenever possible to reduce repetition of basic operations already covered in the construction of $\mathsf{VECK_{El}}$.
Towards this end, we split the operations $\mathsf{VECK_{El}.Enc}\rightarrow \mathsf{(vk, sk, ct, \pi)}$ into two sub-operations:
\[
\mathsf{VECK_{El}.Enc_1}\rightarrow \mathsf{(vk, sk, ct)}
\quad\text{and}\quad
\mathsf{VECK_{El}.Enc_2}\rightarrow (\pi, \mathsf{ct}_{-}).
\]
The first sub-operation returns all outputs except the proof $\pi$, while the second sub-operation returns only the proof along with $\mathsf{ct}_{-}$, an encryption at point $-1$ employed in $\mathsf{VECK_{El}.Enc}$ to assist with ciphertext verification.

Furthermore, we stipulate that if $\mathsf{VECK_{El}.Dec}$ fails to recover a particular value (i.e., the plaintext lies outside the brute-force search range), it returns $\bot$ \emph{for that specific position only}, which we interpret as an erasure.

We employ RS detection and decoding as black-box subroutines defined as follows:
\[
\mathsf{RS.Det}(S,\{\tilde{\phi}_i\}_{i\in S}) \;\rightarrow\; \{0,1\}
\quad\text{and}\quad
\mathsf{RS.Dec}(S',\,S,\,\{\tilde{\phi}_i\}_{i\in S}) 
\;\rightarrow\; \{\phi(i)\}_{i\in S'}\cup\{\bot\},
\]
where $S\subseteq\mathbb{F}_p$ is an ordered set of evaluation points, $S'$ is the target set (typically a subset of $S$), and $\{\tilde{\phi}_i\}_{i\in S}$ denotes the ordered sequence of received symbols, which may potentially be corrupted.

Specifically, $\mathsf{RS.Det}$ checks whether the received sequence is a valid RS codeword, returning $0$ if it is valid and $1$ otherwise.
The decoding algorithm $\mathsf{RS.Dec}$ takes the (potentially corrupted) received sequence and attempts to reconstruct the polynomial evaluations at the points in the subset $S'$. If decoding succeeds, it outputs the corrected evaluations $\{\phi(i)\}_{i\in S'}$; otherwise, it returns $\bot$.

\subsection{FDE via $\VECKplus$}
Recall that a polynomial $\phi(X) \in \F_p[X]$ of degree at most $\ell$ is
uniquely determined by any $m{+}1 \ge \ell{+}1$ distinct evaluations, and
that RS decoding still recovers $\phi$ even if up to
$\big\lfloor (m-\ell)/2 \big\rfloor$ of these evaluations are arbitrarily
corrupted at unknown positions.
We exploit this redundancy by checking only a capped number $R$ of
pseudorandomly chosen positions.
If $m$ is chosen sufficiently larger than $\ell$, then, except with
probability at most $\epsilon$, this limited check guarantees that the
number of corrupt evaluations is within the RS decoding radius, thus $\phi$
(and hence the underlying file) remains recoverable.

\color{black}
The following lemma quantifies this trade-off by giving, for a sample size
$R$ and target failure probability $\epsilon = 2^{-\lambda}$, a corresponding
lower bound $\beta_{\min}(\lambda,R)$ on the redundancy factor $\beta$ (and
hence on $m = \lceil \beta(\ell{+}1)\rceil$).
\color{black}

\begin{lemma}[Redundancy vs.\ sample size]\label{lem:hitting}
Let \(k:=\ell{+}1\) and \(m=\lceil \beta k\rceil\) for some \(\beta>1\).
Let \(C\subseteq[m]\) be any set of \emph{corrupted positions} with \(|C|>t:=\big\lfloor (m-k)/2\big\rfloor\).
Let \(S_R\subseteq[m]\) be a uniformly random subset of size~\(R\).
Then
\[
\Pr\!\big[S_R\cap C=\emptyset\big]\;\le\;\left(\frac{\beta+1}{2\beta}\right)^{R}.
\]
In particular, to ensure \(\Pr\!\big[S_R\cap C=\emptyset\big]\le 2^{-\lambda}\) it suffices to choose
\[
R \;\ge\; \left\lceil \frac{\lambda}{\log_2\!\Big(\tfrac{2\beta}{\beta+1}\Big)} \right\rceil
\quad\Longleftrightarrow\quad
\beta \;\ge\; \beta_{\min}(\lambda,R) \;:=\; \frac{2^{\lambda/R}}{\,2-2^{\lambda/R}\,}.
\]
\end{lemma}

\begin{proof}
Let \(\delta:=|C|/m\).
Since \(|C|>t=\lfloor(m-k)/2\rfloor\), we have
\[
\delta \;\ge\; \frac{t+1}{m} \;\ge\; \frac{m-k+1}{2m}
\;\ge\; \frac{\beta-1}{2\beta}
\quad\text{(using }m=\lceil\beta k\rceil\text{)}.
\]
\color{black}
For sampling, observe that $S_R$ is obtained by choosing $R$ indices
without replacement from the $m$ positions, of which $|C|=\delta m$
are corrupted. Thus the number of corrupted indices in $S_R$ follows
a hypergeometric distribution $\mathrm{Hypergeom}(N=m,K=|C|,n=R)$,
and in particular
\[
  \Pr[S_R \cap C = \varnothing]
   = \Pr[X=0]
   = \frac{\binom{m-|C|}{R}}{\binom{m}{R}}
   \le (1-\delta)^R,
\]
where the inequality uses
$\prod_{j=0}^{R-1}\frac{m-|C|-j}{m-j} \le
\bigl(\frac{m-|C|}{m}\bigr)^R = (1-\delta)^R$.
\color{black}
Plugging \(\delta\ge(\beta-1)/(2\beta)\) gives
\(\Pr[S_R\cap C=\emptyset]\le \big((\beta+1)/(2\beta)\big)^R\).
To make this at most \(2^{-\lambda}\), it suffices that
\(\big((\beta+1)/(2\beta)\big)^R\le 2^{-\lambda}\),
which rearranges to
\(R \ge \lambda/\log_2\!\big(2\beta/(\beta+1)\big)\).
Equivalently, for a given \((\lambda,R)\),
\(\beta \ge \beta_{\min}(\lambda,R):=2^{\lambda/R}/\bigl(2-2^{\lambda/R}\bigr)\).
\end{proof}

\begin{example}
By Lemma~\ref{lem:hitting}, the minimal redundancy factor that suffices for a given security parameter \(\lambda\) and sample budget \(R\) is
\[
\beta_{\min}(\lambda, R)\;=\;\frac{2^{\lambda/R}}{\,2-2^{\lambda/R}\,}.
\]
Note that \(\beta_{\min}(\lambda,R)\) is strictly decreasing in \(R\).\\
For \(\lambda=128\) and assuming \(\ell{+}1>R\), plugging in \(R\in\{512,1024\}\) gives
$\beta_{\min}(128,512)\approx 1.467\ \ (\text{overhead }46.7\%)$ and
$\beta_{\min}(128,1024)\approx 1.199\ \ (\text{overhead }19.9\%)$.

\end{example}

We introduce a \emph{sample budget} \(R\in\mathbb{N}\) and set the verifier’s sample size to
\[
\lvert S_R\rvert \;=\; \min(\ell{+}1,\,R).
\]
We then choose the RS redundancy (equivalently, pick \(\beta>1\) and set \(m=\lceil \beta(\ell{+}1)\rceil\)) so that a uniform sample of size \(\lvert S_R\rvert\) intersects any corruption set larger than the RS correction radius except with probability at most \(2^{-\lambda}\). By Lemma~\ref{lem:hitting}, it suffices to ensure
\[
R \;\ge\; \left\lceil \frac{\lambda}{\log_2\!\Big(\tfrac{2\beta}{\beta+1}\Big)} \right\rceil
\quad\Longleftrightarrow\quad
\beta \;\ge\; \beta_{\min}(\lambda,R) \;:=\; \frac{2^{\lambda/R}}{\,2-2^{\lambda/R}\,}.
\]
(For a finite \(\beta\), this requires \(R>\lambda\).) Operationally:
\begin{itemize}
  \item If \(\ell{+}1\le R\): take \(\beta=1\) and \(m=\ell{+}1\) (no RS extension) and verify all \(\ell{+}1\) positions.
  \item If \(\ell{+}1>R\): fix \(\lvert S_R\rvert=R\), choose any \(\beta\ge \beta_{\min}(\lambda,R)\), and set \(m=\lceil \beta(\ell{+}1)\rceil\).
\end{itemize}
This caps the verifier’s cost by \(R\) while letting redundancy \(\beta\) be chosen to meet a target soundness \(2^{-\lambda}\). 

\paragraph{Protocol outline.}
Given \(m\) as above, the server computes \emph{exponential ElGamal} encryptions of the \(m\) RS-extended evaluations and sends all \(m\) ciphertexts to the client. For correctness, it proves only a Fiat-Shamir-derived subset \(S_R\subseteq [m]\) of size \(\lvert S_R\rvert=\min(\ell{+}1,R)\), linking those sampled ciphertexts to the original KZG commitment. The client verifies these \(\lvert S_R\rvert\) checks (cost \(O(\lvert S_R\rvert)\le O(R)\)) and, after payment, decrypts all \(m\) symbols. A lightweight RS syndrome test detects any corruption; only if needed, RS decoding recovers the original \(\ell{+}1\) evaluations. Except with probability at most \(2^{-\lambda}\), any deviation that exceeds the RS correction radius is caught by the sample, so client verification is independent of the file size.

In what follows we present $\VECKplus$ for the full-retrieval setting, where the client requests the entire dataset (i.e., \(S=[\ell]\)). 
The subset-retrieval case \(S \subset [\ell]\)---which employs the same RS-based technique---is deferred to Appendix~\ref{app:subset}.

When the sample budget satisfies \(R \ge \ell{+}1\), $\VECKplus$ collapses to the ElGamal-based \(\mathsf{VECK_{EL}}\) of Tas et al.---no RS redundancy is needed (\(\beta{=}1,\ m{=}\ell{+}1\)) and all \(\ell{+}1\) positions are verified---so we focus on the non-trivial regime \(R<\ell{+}1\).

\color{black}
\paragraph{Fiat--Shamir subset derivation.}
Given a hash function $H$ modeled as a random oracle, we define
\[
  S_R \;\gets\; \mathsf{DeriveSubset}_H(C,\mathsf{vk},\mathsf{ct},m,R)
\]
as a deterministic procedure that derives an $R$-element subset
$S_R \subseteq [m]$ from the transcript $(C,\mathsf{vk},\mathsf{ct})$.
For concreteness, one may implement $\mathsf{DeriveSubset}_H$ as follows.
For $j = 1,2,\ldots$ query
\[
  \sigma_j \;\coloneqq\; H(C,\mathsf{vk},\mathsf{ct},m,R,j) \in \{0,1\}^{\lambda'}
\]
and interpret $\sigma_j$ as a candidate index in $[m]$ using standard
rejection sampling (discarding values outside $[m]$ and duplicates)
until $R$ distinct indices have been collected; their set is output as $S_R$.
\color{black}

\noindent

\paragraph{\textbfmath{$\mathsf{\VECKplus}$: case $S=[\ell]$ (client requests the entire data set).}}
\begin{itemize}
  \item $\mathsf{\VECKplus.GEN}(\mathsf{crs}) \rightarrow \mathsf{pp}$: 
  On input $\mathsf{crs}=\bigl(\mathcal{G},\{g_1^{\tau^i}\}_{i=0}^{n},\{g_2^{\tau^i}\}_{i=0}^{n}\bigr)$, sample $h\!\leftarrow_R\! G_1$ and $h_i\!\leftarrow_R\! G_1$ for all $i\in [m]\cup\{-1\}$, where $m=\lceil \beta(\ell+1)\rceil$. 
  The values of $m$ and $\beta$ are derived from the sample budget $R$ via Lemma~\ref{lem:hitting}.

\item[] 
\item $\mathsf{\VECKplus.Enc}(F_{[\ell]}, C_\phi, \phi(X))
        \rightarrow \mathsf{(vk, sk, ct, \pi)}$.
\begin{enumerate}
  \item Compute $\bigl(\mathsf{vk},\,\mathsf{sk},\,\mathsf{ct}\bigr) \coloneqq 
        \mathsf{VECK}_{\mathsf{El}}.\mathsf{Enc}_1\!\bigl(F_{[m]},\,C_\phi,\,\phi(X)\bigr)$.
  \item Derive a Fiat-Shamir challenge subset $S_R \subseteq [m]$ of size 
        $\lvert S_R\rvert= R$ deterministically from the transcript, e.g.,
        \[
          S_R \;\gets\; \mathsf{DeriveSubset}_H\bigl(C_\phi,\,\mathsf{vk},\,\mathsf{ct},\,m,\,R \bigr).
        \]
  \item Compute 
        \(
          \bigl(\pi_R,\,\mathsf{ct}_{-}\bigr) \coloneqq 
          \mathsf{VECK}_{\mathsf{El}}.\mathsf{Enc}_2\!\bigl(F_{S_R},\,C_\phi,\,\phi(X)\bigr).
        \)
  \item Output $\bigl(\mathsf{vk},\,\mathsf{sk},\,\mathsf{ct},\,\pi\bigr)$ with 
        $\pi \coloneqq \bigl(\pi_R,\,\mathsf{ct}_{-}\bigr)$.
\end{enumerate}

    \item[]
    \item $\mathsf{\VECKplus.Ver_{ct}(F_{[\ell]}, C_\phi, vk, ct, \pi)} \rightarrow 0/1$
    \begin{enumerate}
        \item Parse $\pi$ as $(\pi_R,  \mathsf{ct}_{-})$.
        \item Output $\mathsf{VECK_{El}.Ver_{ct}(F_{S_R}, C_\phi, vk, \{ct_i\}_{i\in S_R}\cup ct_{-}, \pi_R)}$.
    \end{enumerate}

    \item[]
    \item $\mathsf{\VECKplus.Ver_{key} (vk,sk)} 
    \rightarrow 0/1:$ For $\mathsf{sk}=s\in \mathbb{F}_p$, return 1 iff $\mathsf{vk}=h^s$.
    
    \item[]
    \item $\mathsf{\VECKplus.Dec(F_{[m]}, sk, ct)}
        \rightarrow \{\phi(i)\}_{i\in [\ell]}$.
    \begin{enumerate}
        \item Compute $\{\tilde{\phi}(i)\}_{i\in[m]}\coloneqq \mathsf{VECK.Dec(F_{[m]}, sk, ct)}$.
        \item If $\mathsf{RS.Det}([m],\{\tilde{\phi}(i)\}_{i\in[m]}) = 0$ (no error detected), then output $\{\tilde{\phi}(i)\}_{i\in[\ell]}$. Otherwise, run the decoding algorithm and output
        \[
        \mathsf{RS.Dec}\left([\ell], [m], \{\tilde{\phi}(i)\}_{i\in[m]}\right).
        \]
    \end{enumerate}

\end{itemize}

\begin{theorem}[Security of $\VECKplus$]\label{thm:veckplus-security}
Fix a security parameter \(\lambda\) and a sample budget \(R\) with \(\lambda < R < \ell{+}1\).
Let
\[
\beta_{\min}(\lambda,R)\;:=\;\frac{2^{\lambda/R}}{\,2-2^{\lambda/R}\,}
\]
and choose any redundancy \(\beta \ge \beta_{\min}(\lambda,R)\), setting \(m=\lceil \beta(\ell+1)\rceil\).
Let \(S_R\subseteq[m]\) be the Fiat-Shamir-derived challenge subset with \(|S_R|=R\).
In the random-oracle and algebraic-group models, \(\VECKplus\) satisfies correctness,
\emph{soundness with negligible failure probability in \(\lambda\)}, and computational
zero-knowledge; hence \(\VECKplus\) is a secure \(\mathsf{VECK}\).
\end{theorem}

\begin{proof}
    See Appendix~\ref{app:veckplus-security} 
\end{proof}

\begin{theorem}[Verification cost for full retrieval]
\label{thm:veckplus-ver-cost}
Let $S=[\ell]$, and assume the verifier derives the Fiat--Shamir challenge by
incrementally hashing the ciphertext transcript as it is received. Then, the client-side verification
$\VECKplus.\mathsf{Ver}_{ct}$ runs in $O(\lambda)$ time, independent of~$\ell$.
\end{theorem}

\begin{proof}
In the case $S=[\ell]$, $\VECKplus.\mathsf{Ver}_{ct}$ parses the proof and invokes
$\mathsf{VECK}_{\mathsf{El}}.\mathsf{Ver}_{ct}$ on a Fiat--Shamir-derived challenge subset $S_R$
of size $\lvert S_R\rvert=\Theta(\lambda)$, together with a constant amount of auxiliary data.
The verifier $\mathsf{VECK}_{\mathsf{El}}.\mathsf{Ver}_{ct}$ performs a number of checks
(linear combinations/MSMs and a constant number of pairings) that is linear in $\lvert S_R\rvert$,
and all remaining work is constant. Therefore the total running time is
$\mathcal{O}(\lvert S_R\rvert)=\mathcal{O}(\lambda)$.
\end{proof}

\subsection{FDE via $\VECKstar$}
\color{black}
While $\VECKplus$ already caps the verifier’s work at a small constant once the file exceeds the sample budget, it still encrypts the \emph{entire} file under exponential ElGamal, leading to a $>\!10\times$ bandwidth blowup and non-trivial client-side decryption costs. To address these remaining bottlenecks, we introduce a second instantiation, $\VECKstar$.

Building on the RS-based approach, $\VECKstar$ adopts a \emph{two-layer} design. $\VECKstar$ first encodes with constant-rate RS and masks the entire extended codeword using a fast, hash-derived stream (symmetric). It then select a pseudorandom (Fiat--Shamir-derived) $\Theta(\lambda)$ subset of positions and encrypt \emph{only} those with exponential ElGamal. A file-size-independent zk-SNARK proves, for the sampled positions, that the masked values and the ElGamal plaintexts are consistent under the same secret key; \emph{outside} the SNARK, it runs $\mathsf{VECK}_{\mathsf{El}}.\mathsf{Ver}_{ct}$ to check that the sampled ElGamal ciphertexts are consistent with the original KZG commitment. This two-step consistency (as opposed to a single-step SNARK-based consistency check) minimizes the expensive SNARK work and improves the overall performance.
\color{black}


As a result, both encryption and decryption are symmetric-key fast, and the total communication overhead falls below \(50\%\) of the plaintext for \(R\!\ge\!512\), compared to the \(\sim 10\times\) overhead in the original FDE implementations. The trade-off is the zk-SNARK cost: for \(R = 512\), for example, SNARK proving takes about \(10\,\mathrm{s}\) and verification about \(0.1\,\mathrm{s}\) on a 16-core machine (Figure~\ref{fig:VECKstar}).

\noindent 
\paragraph{\textbfmath{$\mathsf{\VECKstar}$: case $S=[\ell]$ (client requests the entire data set).}}
    \begin{itemize}
      \item $\mathsf{\VECKstar.GEN(crs)} \rightarrow \mathsf{pp}$:
      On input $\mathsf{crs}=\bigl(\mathcal{G},\{g_1^{\tau^i}\}_{i=0}^{n},\{g_2^{\tau^i}\}_{i=0}^{n}\bigr)$\footnote{Here $n$ may be taken as small as $R$.}, sample random group elements with unknown discrete logarithms $h \gets_R G_1$ and $h_i \gets_R G_1$ for $i \in [m]\cup\{-1\}$, where $m := \lceil \beta(\ell+1)\rceil$.

    \item $\mathsf{\VECKstar.Enc(F_{[\ell]}, C_\phi, \phi(X)})
        \rightarrow \mathsf{(vk, sk, ct, \pi)}$.
    \begin{enumerate}
        \item Sample $s \gets_{R} \mathbb{F}_p$, set $\mathsf{sk} := s$, $\mathsf{vk} := h^s$.
        
        \item Compute $\mathsf{ct}=\{\mathsf{ct}_i\coloneqq \phi(i)+H(\mathsf{sk}, i)\}_{i\in [m]}$.

        \item Generate the Fiat-Shamir challenge
          \[
            S_R \;=\;
            \mathsf{DeriveSubset}_H\bigl(C_\phi,\,\mathsf{vk},\,\mathsf{ct},
              m,\;R\bigr) \bigr), \quad \text{where }|S_R|=R.
          \]

        \item Compute 
        $(\text{--},\,\text{--},\,\mathsf{ct}',\,\pi_R)
        \;\coloneqq\;
        \mathsf{VECK_{El}.Enc}\bigl(F_{S_R},\,C_\phi,\,\phi(X)\bigr)$ --- ignore the $(\mathsf{vk},\mathsf{sk})$ returned by \texttt{VECK\_El.Enc}
         
        \item Compute a zk-SNARK proof $\pi_Z$ for the relation $\mathcal{R}$:
        \[
            \mathcal{R} \;=\;
            \left\{
            \begin{array}{l}
            \bigl(\mathsf{vk},\,h,\,\{h_i\}_{i\in S_R},\,\\
            \{\mathsf{ct}_i\}_{i\in S_R},\,\{\mathsf{ct}'_i\}_{i\in S_R}\bigr),\\
            \bigl(\mathsf{sk},\,\{x_i\}_{i\in S_R}\bigr)
            \end{array}
            \;\middle|\;
            \begin{array}{l}
            \mathsf{vk} = h^{\mathsf{sk}},\\[4pt]
            \forall i\in S_R:\; \mathsf{ct}_i   = x_i + H(\mathsf{sk},i),\\[4pt]
            \forall i\in S_R:\; \mathsf{ct}'_i = h_i^{\,\mathsf{sk}}\,g_1^{x_i}
            \end{array}
            \right\}.
        \]

        \item Output $\mathsf{(vk, sk, ct, \pi=
        \mathsf{(\pi_Z, \pi_R, ct')}})$        
    \end{enumerate}

    \item[]
    \item $\mathsf{\VECKstar.Ver_{ct}(F_{[\ell]}, C_\phi, vk, ct, \pi)} \rightarrow 0/1$
    \begin{enumerate}
        \item Parse $\pi$ as $(\pi_Z, \pi_R,  \mathsf{ct'})$.
        \item Compute $b_1 \coloneqq \mathsf{VECK_{El}.Ver_{ct}(F_{S_R}, C_\phi, vk, ct', \pi_R)}$.
        \item Compute $b_2\coloneqq \mathsf{Snark.Ver}(\pi_Z)$.
        \item Output $b_1\wedge b_2$.
    \end{enumerate}

    \item[]
    \item $\mathsf{\VECKstar.Ver_{key} (vk,sk)} 
    \rightarrow 0/1:$ For $\mathsf{sk}=s\in \mathbb{F}_p$, return 1 iff $\mathsf{vk}=h^s$.
    
    \item[]
    \item $\mathsf{\VECKstar.Dec(F_{[m]}, sk, ct)}\rightarrow \{\phi(i)\}_{i\in[\ell]}$
    \begin{enumerate}
        \item Compute the unmasked evaluations $\{\tilde{\phi}(i)\coloneqq \mathsf{ct}_i - H(\mathsf{sk}, i)\}_{i\in [m]}$.
        \item If $\mathsf{RS.Det}([m],\{\tilde{\phi}(i)\}_{i\in[m]}) = 0$ (no error detected), then output $\{\tilde{\phi}(i)\}_{i\in[\ell]}$. Otherwise, run the decoding algorithm and output
        \[
        \mathsf{RS.Dec}\left([\ell], [m], \{\tilde{\phi}(i)\}_{i\in[m]}\right).
        \]
    \end{enumerate}

\end{itemize}

\begin{theorem}[Security of $\VECKstar$]
\label{thm:veckstar-security}
Fix a security parameter $\lambda$ and a sample budget $R$ with $\lambda<R$. 
Choose any redundancy $\beta\ge\beta_{\min}(\lambda,R)$, set $m=\lceil\beta(\ell+1)\rceil$, and let the Fiat--Shamir challenge subset $S_R\subseteq[m]$ satisfy $|S_R|=R$.
Then, in the random-oracle model and the algebraic-group model, $\VECKstar$ satisfies correctness, \emph{soundness with negligible failure probability in~$\lambda$}, and computational zero-knowledge; hence $\VECKstar$ is a secure $\mathsf{VECK}$.
\end{theorem}
\begin{proof}
    See Appendix~\ref{app:veckstar-security} 
\end{proof}


\begin{theorem}[Cost of $\VECKstar$ for full retrieval]
\label{thm:veckstar-cost}
Let $S=[\ell]$ and let the Fiat-Shamir challenge select a sample $S_R$ with $|S_R|=R=\Theta(\lambda)$.
Then in $\VECKstar$, the \emph{verifier’s time} is $O(\lambda)$---independent of $\ell$.
\end{theorem}

\begin{proof}
Verification performs two checks:  
(i) $\mathsf{VECK_{El}.Ver_{ct}}$ on the $R$ sampled positions, which is linear in $|S_R|$, hence $O(\lambda)$; and  
(ii) $\mathsf{Snark.Ver}(\pi_Z)$, which is constant time for succinct SNARKs whose running time is a linear function of $|S_R|$ and is independent of $\ell$
Therefore, the verifier’s time is $O(\lambda)$.  

\end{proof}

\noindent Figure~\ref{fig:VECKstar} reports proving/verification times for $R\in\{256,512,1024\}$ on a multi‑core CPU (Intel Core i9‑13900KF, 32\,GiB RAM). Using 8 cores and $R=512$ (corresponding to $<\!50\%$ bandwidth overhead), proving takes about $10$\,s and verification about $0.1$\,s. Notably, unlike the ElGamal‑based baseline, $\VECKstar$ requires no range proofs for ElGamal plaintexts---the sampled ElGamal ciphertexts are never decrypted---removing a major prover‑side cost. Groth16 proofs over BW6\mbox{-}761 are constant‑size ($\approx\!288$\,B), and both verification time and proof size are effectively independent of the file size (they depend only on $R$).

\section{FDE via Payment Channels}

\color{black}
Tas et al.~\cite{TasSZMKBN24} proposed a Bitcoin-based implementation of FDE that uses adaptor signatures to tie the VECK decryption key to a payment. However, their concrete VECK instantiations live on a pairing-friendly curve (e.g., BLS12-381), whereas Bitcoin signatures and adaptor signatures are standardized on secp256k1. This curve mismatch means their Bitcoin construction is, at present, only a high-level template: it cannot be deployed with the primitives available on today’s Bitcoin network.

In this section we show how to make the VECK-based FDE protocol deployable on Bitcoin today. Our construction runs the payment leg over standard Lightning Network (LN) hash time-locked contracts (HTLCs) and uses a small zk-SNARK, in the spirit of Zero-Knowledge Contingent Payments (ZKCP)~\cite{MaxwellZKCP}, to cryptographically tie the BLS12-381 VECK decryption key to the secp256k1 secret used in the HTLC. This both resolves the curve mismatch and makes the common case entirely off-chain, while preserving the correctness and fairness guarantees of FDE.

\color{black}

Specifically, the server (prover) generates a zk-SNARK proof $\pi_t$ attesting knowledge of a secret scalar $\mathsf{sk}$ that satisfies the following relation:
\[
R\bigl((h, \mathsf{vk}, t), \mathsf{sk}\bigr) = [ \mathsf{vk} = h^{\mathsf{sk}} \wedge t = H(\mathsf{sk})],
\]
where $h$, $\mathsf{vk}$, and $t$ are publicly known parameters. Although generating the zk-SNARK proof $\pi_t$ introduces computational overhead, this cost is constant with respect to the data size and can be precomputed.

Our resulting FDE protocol operates in two steps. In step 1, The server sends to the client the ciphertext and its proof, the value $t$, and the zk-SNARK proof $\pi_t$. 
In step 2, the client verifies all provided proofs, including $\pi_t$. Upon successful verification, the client initiates a payment via LN, using $t$ as the hash value in the HTLC.
Due to the atomicity provided by LN, the server obtains the payment \emph{iff} it reveals the secret scalar~$\mathsf{sk}$, enabling the client to decrypt the ciphertext. Crucially, this achieves atomic exchange without incurring additional costly blockchain transactions.

\section{Evaluation}
\color{black}

\begin{figure}[h]
    \centering
    \includegraphics[width=0.8\linewidth]{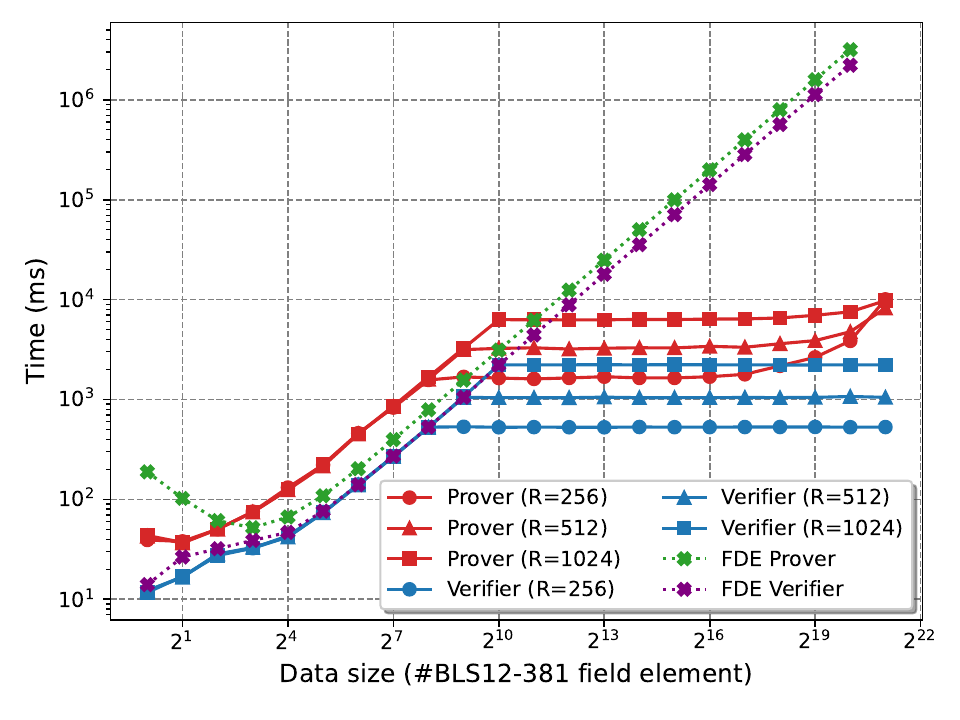}
    \vspace{-16pt}
    \caption{\textcolor{black}{Proof verification and generation times in $\VECKplus$ and $\VECK$.}}
    \label{fig:VECKplus}
\end{figure}

We implemented $\VECKplus$ by forking the open-source FDE code of Tas et al.~\cite{TasSZMKBN24}\footnote{\color{black}Our modified code is available at \url{https://github.com/SecureX-UofA/fde-plus}.\color{black}} and replacing their ElGamal-based $\mathsf{VECK_{El}}$ backend with our RS-sampled variant.
We then evaluated both $\VECKplus$ and $\mathsf{VECK_{El}}$ on a commodity desktop (Intel Core i9-13900KF, 32\,GiB RAM).
Figure~\ref{fig:VECKplus} plots verifier and prover time as a function of file size.
As predicted by Theorem~\ref{thm:veckplus-ver-cost}, the $\VECKplus$ verifier curve flattens once the file size meets the sample budget (i.e., when $\ell+1 \ge R$); for $R=512$, it stabilizes at roughly $1$\,s, capping client-side verification at about $1$\,s on this machine for any file size.
By contrast, both proving and verification for $\mathsf{VECK_{El}}$ grow essentially linearly with the file size.

Figure~\ref{fig:VECKplus} also shows the proof-generation time for $\VECKplus$.
Unlike verification, proving scales with the file size.\footnote{Although the proof touches only the $R$ sampled indices $S_R$, the prover must interpolate a polynomial of degree at most $R{+}1$ and prove its consistency with the commitment; this consistency step grows linearly with $\ell$, not with $R$.}
Even so, it substantially improves over existing FDE implementations: on our hardware, the ElGamal baseline requires $\approx 6{,}295\,\mathrm{s}$ to generate a proof for a $32\,\mathrm{MiB}$ file, whereas with $R=512$ our prover completes in $\approx 4.8\,\mathrm{s}$—an $\sim\!1{,}300\times$ speedup.

\vspace{-10pt}
\vspace{-10pt}


$\VECKstar$ replaces ElGamal over the full file with a fast, hash-derived mask and confines public-key work to a $\Theta(\lambda)$ sample. As a result, encryption and decryption are symmetric-key fast and total communication drops to about $1.5\times$ the plaintext (i.e., $<\!50\%$ overhead) by setting $R\geq 512$, down from $\gtrsim 10\times$ all at the file-size independent cost of proving and verifying $\pi_Z$ (Step~5 of $\mathsf{VECK^\star.Enc}$)

To assess the cost of proving and verifying $\pi_Z$, we implemented the circuit in \texttt{gnark} and used Groth16. For performance, we instantiate the statement group as $G_1(\mathrm{BLS12\text{-}377})$ and prove over the matching BW6\mbox{-}761 outer curve (a standard 2\mbox{-}chain), thus elliptic-curve operations are \emph{native} in the circuit. We use a SNARK‑friendly pseudorandom function (PRF)---MiMC (Minimal Multiplicative Complexity)---for $H(\mathsf{sk},i)$ and enforce range checks $0\le \mathsf{sk},x_i<r_{377}$.
 An implementation that keeps $G_1$ inside the SNARK is left to future work.

\begin{figure}
    \centering
    \includegraphics[width=0.76\linewidth]{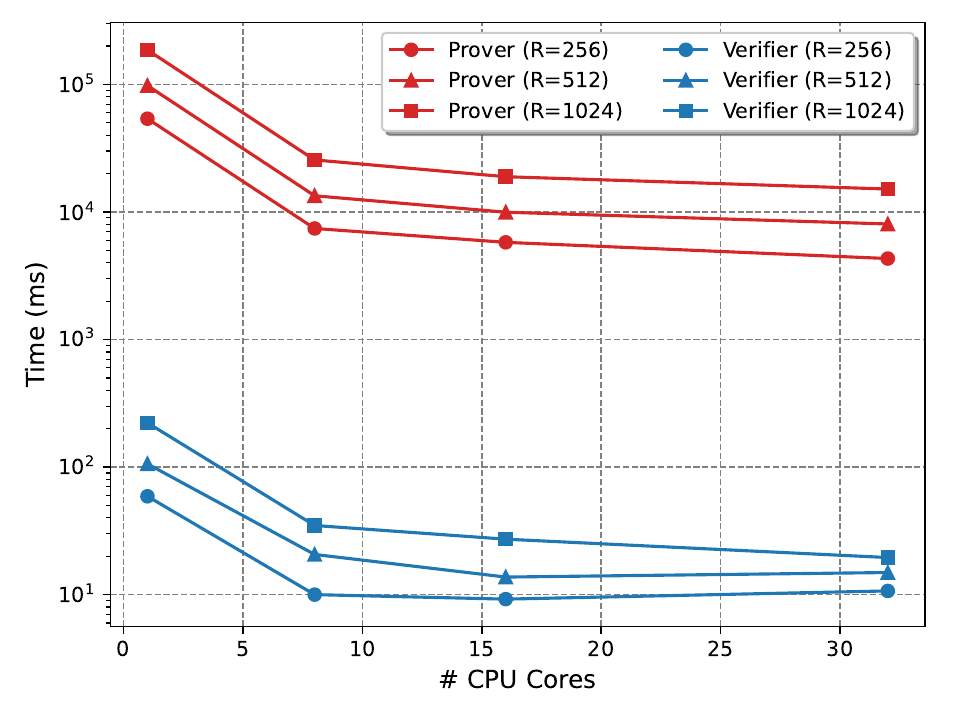}
    \vspace{-10pt}
    \caption{SNARK proof ($\pi_Z$) generation and verification times.}
    \label{fig:VECKstar}
\end{figure}

Figure~\ref{fig:VECKstar} reports proving/verification times for $R\in\{256,512,1024\}$ on a multi‑core CPU (Intel Core i9‑13900KF, 32\,GiB RAM). Using 8 cores and $R=512$ (corresponding to $<\!50\%$ bandwidth overhead), proving takes about $10$\,s and verification about $0.1$\,s. Notably, unlike the ElGamal‑based baseline, $\VECKstar$ requires no range proofs for ElGamal plaintexts---the sampled ElGamal ciphertexts are never decrypted---removing a major prover‑side cost. Groth16 proofs over BW6\mbox{-}761 are constant‑size ($\approx\!288$\,B), and both verification time and proof size are effectively independent of the file size (they depend only on $R$).

\vspace{-15pt}
\vspace{-15pt}

\section{Conclusion}
We revisited FDE and showed that prover and verifier time, as well as bandwidth overhead, can be drastically reduced. Our first design, $\mathrm{VECK}^{+}_{\mathrm{EL}}$, pairs KZG commitments with Reed-Solomon sampling to cap client verification at $\mathcal{O}(\lambda)$ regardless of data length; in our prototype it plateaus at $\approx 1\,\mathrm{s}$ once the sample budget is met, while prover time drops by orders of magnitude (e.g., $32\,\mathrm{MiB}:~6295\,\mathrm{s}\!\rightarrow\!4.8\,\mathrm{s}$). Our second design, $\mathrm{VECK}^{\star}_{\mathrm{EL}}$, removes the remaining bottlenecks by masking the full file with a hash‑derived one‑time stream and confining public‑key work to a $\Theta(\lambda)$ sample tied together with a file‑size-independent zk‑SNARK. This makes encryption/decryption symmetric-key fast, reduces communication from $\gtrsim 10\times$ plaintext to $<1.5\times$ (i.e., $<50\%$ overhead), and adds only $<0.1\,\mathrm{s}$ to verification and $\approx 10\,\mathrm{s}$ of proof generation on 8-16 cores. Finally, a compact SNARK bridge binds the committed decryption key to a Lightning HTLC hash, enabling fully off‑chain FDE with fees $<\$0.01$ and seconds‑scale latency while preserving FDE’s constant on‑chain footprint and fairness guarantees. Taken together, these results move FDE from promising blueprint to deployable primitive: verification is essentially constant in $\lambda$, bandwidth is near-plaintext, and the remaining bottlenecks are network and storage rather than cryptography—opening a practical path to fair, atomic, pay‑per‑file exchange for large datasets at near network speed.

\clearpage
\appendix

\renewcommand{\theHsection}{A\arabic{section}}
\section*{\Large Appendices}

\section{Background on Reed--Solomon Codes}
\label{app:rs-background}

Let $(a_0,\dots,a_\ell)\in\mathbb{F}_p^{\ell+1}$ be a message encoded by the polynomial
$\phi(X)=\sum_{i=0}^{\ell} a_i X^{\,i}$.
Given $n \ge \ell+1$ pairwise-distinct evaluation points
$\alpha_1,\dots,\alpha_n\in\mathbb{F}_p$, the corresponding Reed--Solomon (RS) codeword is
  $\mathbf{c}=(\phi(\alpha_1),\dots,\phi(\alpha_n))\in\mathbb{F}_p^{\,n}$.
This yields an $(n,\ell+1)$ linear code with rate $R=(\ell+1)/n$ and minimum distance
$d_{\min}=n-\ell$, capable of correcting up to
$t=\lfloor(d_{\min}-1)/2\rfloor$ symbol errors.

\smallskip
\noindent\textbf{Fast deterministic detection via the parity check.}\quad
Let $H\in\mathbb{F}_p^{(n-k)\times n}$ be the parity-check matrix of the code
($k=\ell+1$).  
For a received word $\mathbf{c}=(c_1,\dots,c_n)$, compute the syndrome
vector
\[
  \mathbf{s}=H\,\mathbf{c}^{\mathsf T}\in\mathbb{F}_p^{\,n-k}.
\]
This requires only a single $O(n)$ streaming pass.  
If $\mathbf{s}=0$, the word is a valid codeword and full decoding can be skipped; otherwise the decoder is invoked.
This \emph{detect-then-correct} strategy makes the common error-free case inexpensive.

Optionally, one could speed up the detection process by applying the standard random linear-combination (Freivalds) test to compress
the syndrome check to a single scalar:  
Fix a random non-zero vector $v\in\mathbb{F}_p^{\,n-k}$ once and pre-compute $w = v^{\mathsf T}H \in \mathbb{F}_p^{\,n}$.
Then, in one streaming pass, compute
  $T = \sum_{j=1}^{n} w_j\,c_j$.
If $T=0$ we accept immediately; otherwise full decoding is performed.
A non-codeword passes this test with probability $1/p$
(e.g., $2^{-255}$ in the 255-bit BLS12-381 field).

\section{$\VECKplus$ for Subset Retrieval}
\label{app:subset}

Here we present $\VECKplus$ for the subset-retrieval setting \(S \subset [\ell]\).
When the sample budget satisfies \(R \ge |S|\), $\VECKplus$ collapses to the
ElGamal-based \(\mathsf{VECK_{El}}\) of Tas et al.\ (no RS redundancy is needed); we therefore
treat the non-trivial regime \(R<|S|\). The RS redundancy \(\beta>1\) and the
code length \(m=\lceil \beta\,|S|\rceil\) are chosen from the sample budget \(R\) and security
parameter \(\lambda\) as in Lemma~\ref{lem:hitting}, ensuring that a uniform sample
\(S_R\) hits any corruption set beyond the RS correction radius except
with probability at most \(2^{-\lambda}\).  

\paragraph{\textbfmath{\(\mathsf{\VECKplus}\): case \(S\subset[\ell]\) (client requests a subset \(S\)).}}.
\begin{itemize}
  \item \(\mathsf{\VECKplus.GEN}(\mathsf{crs}) \rightarrow \mathsf{pp}\): 
  On input \(\mathsf{crs}=\bigl(\mathcal{G},\{g_1^{\tau^i}\}_{i=0}^{n},\{g_2^{\tau^i}\}_{i=0}^{n}\bigr)\),
  sample \(h\!\leftarrow_R\!G_1\) and \(h_i\!\leftarrow_R\!G_1\) for all \(i\in [m]\cup\{-1\}\),
  where \(m\) and \(\beta\) is chosen from \((\lambda,R)\) via
  Lemma~\ref{lem:hitting}. 
  
  \item[] 
  \item \(\mathsf{\VECKplus.Enc}(F_S, C_\phi, \phi(X)) \rightarrow \mathsf{(vk, sk, ct, \pi)}\).
  \begin{enumerate}
    \item Sample \(t \leftarrow_R \mathbb{F}_p\)
    \item Set \(\phi'_S(X) \coloneqq \phi_S(X) + t\,V_S(X)\).
    \item Compute \(C_S \coloneqq \mathsf{Commit}(\mathsf{crs},\,\phi'_S(X))\).
    \item Compute
          $(\mathsf{vk},\mathsf{sk},\mathsf{ct}) \coloneqq
            \mathsf{VECK_{\!El}.Enc_1}\!\bigl(F_{[m]},\,C_S,\,\phi'_S(X)\bigr).$
          
    \item Compute
          $
            \pi_S \;\coloneqq\; \mathsf{batchOpen}\!\bigl(\mathsf{crs},\;\phi(X)-\phi'_S(X),\;S\bigr).
          $
    \item Derive a Fiat-Shamir challenge subset \(S_R\subseteq [m]\) of size
          \(R\):
          \[
            S_R \leftarrow \mathsf{DeriveSubset}_H\!\bigl(C_S,\,\mathsf{vk},\,\mathsf{ct}, \,m,\,R\bigr).
          \]
    \item Compute         
            $(\pi_R,\,\mathsf{ct}_{-}) \coloneqq
            \mathsf{VECK_{\!El}.Enc_2}\!\bigl(F_{S_R},\,C_S,\,\phi'_S(X)\bigr).$
          
    \item Output \(\bigl(\mathsf{vk},\mathsf{sk},\mathsf{ct},\,\pi\bigr)\) with
          \(\pi \coloneqq (\pi_S,\,\pi_R,\,\mathsf{ct}_{-})\).
  \end{enumerate}

  \item[]
  \item \(\mathsf{\VECKplus.Ver_{ct}}(F_S, C_\phi, \mathsf{vk}, \mathsf{ct}, \pi) \rightarrow 0/1\).
  \begin{enumerate}
    \item Parse \(\pi\) as \((\pi_S,\,\pi_R,\,\mathsf{ct}_{-})\).
    \item Compute
            $b_1 \;\coloneqq\; \mathsf{batchVerify}\bigl(\mathsf{crs},\;C_\phi/C_S,\;S,\;\mathbf{0},\;\pi_S\bigr).$
          
    \item Compute
            $b_2 \;\coloneqq\;
            \mathsf{VECK_{\!El}.Ver_{ct}}\bigl(F_{S_R},\,C_S,\,\mathsf{vk},\,
            \{\mathsf{ct}_i\}_{i\in S_R}\cup \mathsf{ct}_{-},\,\pi_R\bigr).$
          
    \item Output \(b_1 \land b_2\).
  \end{enumerate}

  \item[]
  \item \(\mathsf{\VECKplus.Ver_{key}}(\mathsf{vk},\mathsf{sk}) \rightarrow 0/1\):
        For \(\mathsf{sk}=s\in\mathbb{F}_p\), return \(1\) iff \(\mathsf{vk}=h^{s}\).
  
  \item[]
  \item \(\mathsf{\VECKplus.Dec}(F_{[m]}, \mathsf{sk}, \mathsf{ct}) \rightarrow \{\phi(i)\}_{i\in S}\).
  \begin{enumerate}
    \item Compute the received word on \([m]\):
          \(\{\tilde{\phi}(i)\}_{i\in[m]}\coloneqq
           \mathsf{VECK_{\!El}.Dec}\bigl(F_{[m]}, \mathsf{sk}, \mathsf{ct}\bigr)\).
    \item If \(\mathsf{RS.Det}([m],\{\tilde{\phi}(i)\}_{i\in[m]})=0\) (no error), output
          \(\{\tilde{\phi}(i)\}_{i\in S}\). Otherwise, decode and output
          \[
            \mathsf{RS.Dec}\bigl(S,\,[m],\,\{\tilde{\phi}(i)\}_{i\in[m]}\bigr).
          \]
  \end{enumerate}
\end{itemize}

\section{Proof sketch of Theorem~\ref{thm:veckplus-security}}
\label{app:veckplus-security}
Let $k:=\ell{+}1$, $\beta>1$, $m=\lceil \beta k\rceil$, and let $S_R\subseteq[m]$ be the Fiat-Shamir-derived
challenge subset with $|S_R|=\min(k,R)$. The proof uses (i) polynomial and evaluation binding of KZG, and (ii) the security (soundness and ZK) of the ElGamal-based $\mathsf{VECK}_{\mathsf{El}}$ verifier on $S_R$, as established in the CCS’24 analysis; we merely call that verifier on the sampled positions. 

\paragraph{Correctness.}
In an honest execution, the ciphertexts encrypt the RS-extended evaluations $\{\phi(i)\}_{i\in[m]}$ of the committed polynomial $\phi$,
and the on-sample proof on $S_R$ is accepted by $\mathsf{VECK}_{\mathsf{El}}.\mathsf{Ver}_{ct}$ (completeness).
After payment, decryption yields a valid RS codeword; the syndrome test accepts and, if invoked, decoding recovers the original $k$ evaluations.
Thus the client accepts and outputs the correct values. 

\paragraph{Soundness.}
Let a PPT adversary output $(vk,ct,\pi)$ with $\VECKplus.\mathsf{Ver}_{ct}=1$. 
Let $c\in\mathbb{F}_p^{m}$ be the decrypted vector. Let $t=\lfloor (m-k)/2\rfloor$ be the RS correction radius.

\emph{(a) Few corruptions.} If $c$ differs from the true codeword in at most $t$ positions, RS decoding (recognized by a zero syndrome) returns the correct evaluations.

\emph{(b) Many corruptions.} If $c$ differs from every RS codeword in $e>t$ positions, then the corrupted-position fraction satisfies
\[
\delta=\tfrac{e}{m}\ \ge\ \tfrac{t+1}{m}\ \ge\ \tfrac{m-k+1}{2m}\ \ge\ \tfrac{\beta-1}{2\beta}.
\]
In the ROM, the Fiat-Shamir subset $S_R$ is (to the adversary) indistinguishable from a uniform subset of size $|S_R|$ chosen after $ct$ is fixed.
By Lemma~\ref{lem:hitting}, the miss-probability is at most $(1-\delta)^{|S_R|}\le\bigl(\tfrac{\beta+1}{2\beta}\bigr)^{|S_R|}\le 2^{-\lambda}$ for our choice of $R$.
Conditioned on hitting a corrupted position, acceptance implies a break of either KZG binding or $\mathsf{VECK}_{\mathsf{El}}$ soundness on $S_R$, both assumed secure in ROM/AGM. Hence any accepting transcript is either correct, or breaks an underlying assumption, or occurs with probability at most $2^{-\lambda}$. 

\paragraph{Computational zero-knowledge.}
Define a simulator $\mathsf{Sim}$ on input $C_\phi$. In Hybrid $\mathsf{H}_1$, replace off-sample ciphertexts ($i\notin S_R$) by ElGamal encryptions of random field elements; by DDH these are indistinguishable from encryptions of the true values. In Hybrid $\mathsf{H}_2$, program the random oracle to obtain the desired $S_R$ and replace the on-sample portion $(vk,\{ct_i\}_{i\in S_R},\pi_R)$ with the output of the (computational) ZK simulator for $\mathsf{VECK}_{\mathsf{El}}$ on $S_R$. The hybrids are computationally indistinguishable; Hybrid $\mathsf{H}_2$ equals the simulator’s output. Therefore $(vk,ct,\pi)$ is computationally indistinguishable from real.

\section{Proof sketch of Theorem~\ref{thm:veckstar-security}}
\label{app:veckstar-security}
\color{black}

\textbf{Correctness.} In an honest execution of $\mathsf{\VECKstar.Enc}$, the prover evaluates the committed polynomial $\phi$ on $[m]$ and forms the masked evaluations
\[
  \mathsf{ct}_i \;=\; \phi(i) + H(\mathsf{sk},i)
  \quad\text{for all } i\in[m],
\]
thus the ciphertext vector $\mathsf{ct}$ encodes exactly the RS-extended evaluations of $\phi$ hidden under the hash-derived mask.
On the Fiat–Shamir challenge subset $S_R$, the prover additionally runs $\mathsf{VECK_{El}.Enc}$ on the same evaluations and produces a zk-SNARK $\pi_Z$ certifying that the VECK ciphertexts $\{\mathsf{ct}'_i\}_{i\in S_R}$ and the masked values $\{\mathsf{ct}_i\}_{i\in S_R}$ are consistent with a single key $\mathsf{sk}$ and plaintexts $x_i=\phi(i)$.
By correctness of $\mathsf{VECK_{El}}$ and completeness of the SNARK, the client-side verification $\mathsf{\VECKstar.Ver_{ct}}$ accepts these honestly generated transcripts.
After payment, $\mathsf{\VECKstar.Dec}$ uses the same $\mathsf{sk}$ to remove the mask and then applies $\mathsf{RS.Det}$/$\mathsf{RS.Dec}$; by correctness of the RS code, this recovers the original evaluations $\{\phi(i)\}_{i\in[\ell]}$.
Hence an honest client always accepts and outputs the correct data, hence $\VECKstar$ satisfies correctness.

\textbf{Soundness.}
Let a PPT adversary output $\mathsf{(vk,ct,\pi)}$ with
\[
\mathsf{VECK}^\star_{\mathsf{EL}}.\mathsf{Ver_{ct}}(F_{[\ell]},C_\phi,\mathsf{vk},\mathsf{ct},\pi)=1,
\]
and let $S_R\subseteq [m]$ be the Fiat–Shamir challenge subset determined by $(C_\phi,\mathsf{vk},\mathsf{ct},m,R)$.  Parse $\pi=(\pi_Z,\pi_R,\mathsf{ct}')$, where $\mathsf{ct}'$ are the
ElGamal ciphertexts on $S_R$ and $\pi_R$ is the VECK proof.

By knowledge soundness of the zk-SNARK used in step~5 of
$\mathsf{VECK}^\star_{\mathsf{EL}}.\mathsf{Enc}$ and the fact that
$\mathsf{Snark.Ver}(\pi_Z)=1$, we can extract a secret key $\mathsf{sk}$ and values
$(x_i)_{i\in S_R}$ such that
\[
\mathsf{vk} = h^{\mathsf{sk}},\qquad \mathsf{ct}_i = x_i + H(\mathsf{sk},i),\qquad
\mathsf{ct}'_i = h_i^{\mathsf{sk}}\, g_1^{x_i}\quad\text{for all }i\in S_R.
\]
Thus $(\mathsf{vk},\{\mathsf{ct}'_i\}_{i\in S_R},\pi_R)$ is an accepting transcript for the
ElGamal-based $\mathsf{VECK}_{\mathsf{EL}}$ protocol on the sample $S_R$ with
respect to $C_\phi$.  By the soundness of the ElGamal-based VECK for KZG
commitments \cite{TasSZMKBN24}, we obtain
$x_i = \phi(i)$ for all $i\in S_R$.  Consequently, the unmasked word
$c\in\mathbb{F}_p^m$ defined by
$c_i := ct_i - H(sk,i)$ satisfies $c_i = \phi(i)$ for every $i\in S_R$.

Let $k := \ell+1$ and $t := \lfloor (m-k)/2 \rfloor$ be the RS
correction radius, and let $c^\star\in\mathbb{F}_p^m$ be the RS codeword
corresponding to the true evaluations $\{\phi(i)\}_{i\in[m]}$.
As in Appendix~B, we distinguish two cases.
\emph{(a) Few corruptions.} If $c$ differs from $c^\star$ in at most $t$
positions, then RS decoding in $\mathsf{VECK}^\star_{\mathsf{EL}}.\mathsf{Dec}$
(recognized by a zero syndrome) returns $c^\star$ and hence the correct
evaluations.
\emph{(b) Many corruptions.} Otherwise $c$ differs from every RS codeword in
$e>t$ positions.  Writing $\delta := e/m$, the same calculation as in Lemma~1
yields $\delta \ge (\beta-1)/(2\beta)$, thus a uniformly random subset $S_R$ of
size $R$ hits a corrupted index except with probability at most $2^{-\lambda}$.
In the random-oracle model, the Fiat–Shamir subset derived by
$\mathsf{DeriveSubset}_H(C_\phi,\mathsf{vk},\mathsf{ct},m,R)$ is (to the adversary) indistinguishable from such a random subset once $\mathsf{ct}$ is fixed, and therefore the
miss probability is still at most $2^{-\lambda}$.

Conditioned on $S_R$ hitting a corrupted position, there exists
$j\in S_R$ with $c_j\neq \phi(j)$, contradicting the fact that
$c_j=\phi(j)$ unless we break either KZG binding, the soundness of the
ElGamal-based VECK from~\cite{TasSZMKBN24}, or the soundness of the zk-SNARK.
Hence any accepting transcript in which
$\mathsf{VECK}^\star_{\mathsf{EL}}.\mathsf{Dec}$ outputs an incorrect value
either breaks one of these underlying assumptions or occurs with probability
at most $2^{-\lambda}$.  This yields soundness with negligible failure probability in~$\lambda$.

\textbf{Computational zero-knowledge.}
We show that the view \((\mathsf{vk},\mathsf{ct},\pi)\) of any PPT adversary can be efficiently
simulated from the commitment \(C_\phi\) alone. Define a simulator
\(\mathsf{Sim}\) on input \(C_\phi\) as follows. First sample
\(s \leftarrow \mathbb{F}_p\) and set \(\mathsf{vk} := h^s\). Sample a random vector
\(\mathsf{ct} = (\mathsf{ct}_i)_{i\in[m]} \leftarrow \mathbb{F}_p^m\) and program the random
oracle \(H\) so that (i) for all \(i\in[m]\), the values \(H(s,i)\) are
independent uniform field elements, and (ii) the Fiat–Shamir procedure
\(\mathsf{DeriveSubset}_H(C_\phi,\mathsf{vk},\mathsf{ct},m,R)\) outputs a fixed subset
\(S_R \subseteq [m]\) of size \(R\). Next, invoke the computational
zero-knowledge simulator for the ElGamal-based VECK\(_{\mathrm{El}}\) of
Tas et al.\ on the restriction \(F_{S_R}\) and commitment \(C_\phi\), to
obtain \((\mathsf{ct}',\pi_R)\) distributed as in a real execution on the sampled
positions \(S_R\)~\cite{TasSZMKBN24}. Finally, use the
standard simulator of the zk-SNARK to generate a proof
\(\pi_Z\) for the public statement
\[
(\mathsf{vk},h,\{h_i\}_{i\in S_R},\{\mathsf{ct}_i\}_{i\in S_R},\{\mathsf{ct}'_i\}_{i\in S_R}),
\]
and output \((\mathsf{vk},\mathsf{ct},\pi)\) with \(\pi := (\pi_Z,\pi_R,\mathsf{ct}')\).

Consider hybrids between a real execution of VECK\(^{\star}_{\mathrm{EL}}\)
and the output of \(\mathsf{Sim}\).
In Hybrid~H\(_1\), we keep the ElGamal and SNARK parts as in the real
scheme but view \(H(s,\cdot)\) as a random function keyed by the secret
\(s\). Since \(s\) is never revealed before the decryption phase, the values
\(H(s,i)\) are uniform and independent from the adversary’s perspective, thus
each masked symbol \(\mathsf{ct}_i = \phi(i) + H(s,i)\) is itself uniform in
\(\mathbb{F}_p\); replacing it by an independently uniform \(\mathsf{ct}_i\) does not
change the distribution. In Hybrid~H\(_2\), we replace the real ElGamal
transcript \((\mathsf{ct}',\pi_R)\) on \(S_R\) with the output of the simulator for
VECK\(_{\mathrm{El}}\); by the computational zero-knowledge of
VECK\(_{\mathrm{El}}\), these hybrids are computationally indistinguishable.
In Hybrid~H\(_3\), we replace the real zk-SNARK proof with the simulated
proof for the same public inputs; this is indistinguishable by the
zero-knowledge of the SNARK. The final hybrid is exactly the output of
\(\mathsf{Sim}(C_\phi)\), and each transition incurs only negligible
advantage. Hence \((\mathsf{vk},\mathsf{ct},\pi)\) is computationally indistinguishable from
\(\mathsf{Sim}(C_\phi)\), which depends on \(\phi\) only through the
commitment \(C_\phi\). Therefore VECK\(^{\star}_{\mathrm{EL}}\) satisfies
computational zero-knowledge.

\color{black}

\bibliographystyle{splncs04}
\bibliography{citations}

\end{document}